# Nano-optical imaging of WSe$_2$ waveguide modes revealing light-exciton interactions

Z. Fei[1,2]*, M. E. Scott[3], D. J. Gosztola[1], J. J. Foley IV[1,4], J. Yan[5,6], D. G. Mandrus[5,6], H. Wen[7], P. Zhou[8], D. W. Zhang[8], Y. Sun[1], J. R. Guest[1], S. K. Gray[1], W. Bao[8], G. P. Wiederrecht[1], X. Xu[3,9]

[1]Center for Nanoscale Materials, Argonne National Laboratory, Lemont, Illinois 60439, USA
[2]Department of Physics and Astronomy, Iowa State University, Ames, Iowa 50011, USA
[3]Department of Physics, University of Washington, Seattle, Washington 98195, USA
[4]Department of Chemistry, William Paterson University, Wayne, New Jersey 07470, USA.
[5]Materials Science and Technology Division, Oak Ridge National Laboratory, Oak Ridge, Tennessee 37831, USA
[6]Department of Materials Science and Engineering, University of Tennessee, Knoxville, Tennessee 37996, USA
[7]Advanced Photon Source, Argonne National Laboratory, Lemont, Illinois 60439, USA
[8]State Key Laboratory of ASIC and System, Department of Microelectronics, Fudan University, Shanghai 200433, China
[9]Department of Materials Science and Engineering, University of Washington, Seattle, Washington 98195, USA

*Correspondence to Z. Fei: zfei@iastate.edu

**Abstract**
We report on nano-optical imaging study of WSe$_2$ thin flakes with the scanning near-field optical microscopy (NSOM). The NSOM technique allows us to visualize in real space various waveguide photon modes inside WSe$_2$. By tuning the excitation laser energy, we are able to map the entire dispersion of these waveguide modes both above and below the A exciton energy of WSe$_2$. We found that all the modes interact strongly with WSe$_2$ excitons. The outcome of the interaction is that the observed waveguide modes shift to higher momenta right below the A exciton energy. At higher energies, on the other hand, these modes are strongly damped due to adjacent B excitons or band edge absorptions. The mode-shifting phenomena are consistent with polariton formation in WSe$_2$.

**Main Text**

Group VI transition-metal dichalcogenides (TMDCs) with chemical formula MX$_2$ (M = Mo, W; X = S, Se, Te) are novel semiconductors with layered structures and remarkable (opto)electronic properties [1-4]. Light-matter interactions in this class of materials have been widely explored in the far-field regime where many intriguing optical phenomena due to excitons were observed by emission and reflection/absorption spectroscopy [3-20]. Recently, a few near-field emission studies of TMDCs were reported where nanoscale photoluminescence responses due to grain boundaries were observed [21,22]. Nevertheless, near-field reflection/absorption studies that address finite-momentum light-matter interactions are still missing. Here we perform near-field nano-optical imaging studies of WSe$_2$ thin flakes in the reflection mode by using the aperture-type near-field scanning optical microscopy (NSOM). With this technique, we directly excited finite-momentum waveguide photon modes inside TMDCs and studied their interactions with excitons without the need of additional momentum coupling methods.

The samples studied in our work are thin flakes of tungsten diselenide (WSe$_2$) – a prototypical TMDC [17-19]. The direct excitons of bulk WSe$_2$ are located around 1.57 eV (A excitons) and 2.03 eV (B excitons) corresponding to excitation wavelengths ($\lambda_0$) of about 790 nm and 610 nm, respectively (Fig. S3). As illustrated in Fig. 1a,b, a laser beam is delivered through the NSOM tip with an aperture of ~100 nm in diameter to excite the WSe$_2$ sample underneath the tip. Our laser is a tunable Ti:sapphire oscillator operating in the continuous-wave mode that covers a wide spectral range in the visible and near-infrared regime (700-950 nm). We collect all the reflected or scattered photons through an objective (N.A. = 0.45) above the sample (Fig. 1b) and these photons are counted with an avalanche photodiode (Supplementary Material). Throughout the imaging experiments, we keep the tip and objective fixed and only scan the sample stage. The NSOM collects both the topography and near-field optical images simultaneously at ambient conditions.

The sample investigated in Fig. 1 is a WSe$_2$ flake with a thickness of 260 nm sitting on a standard SiO$_2$/Si wafer (Fig. S1). By tuning $\lambda_0$, we are able to perform spectroscopic nano-imaging of the WSe$_2$ sample. The selected dataset of the near-field images is shown in Fig. 1c-f, where we plot the normalized photon counts. We first consider the near-field image taken at $\lambda_0 = 900$ nm (Fig. 1c). Here, we observe bright fringes on both WSe$_2$ and SiO$_2$ parallel to the sample edge (white dashed line), but the fringe pattern and intensity are clearly different on the two materials. Compared to the fringes on SiO$_2$, those on WSe$_2$ are stronger in intensity and more densely distributed. Moreover, the fringes on WSe$_2$ extend further away from the sample edge and demonstrate no significant damping. As a result, we are able to see fringes (upper right corner) due to a remote edge of WSe$_2$ that is not present within the field of view (Fig. S1b). As $\lambda_0$ decreases (Fig. 1d-f), the intensity of the fringes on WSe$_2$ drops rapidly indicating that their damping becomes significantly larger at shorter $\lambda_0$. Eventually at $\lambda_0 = 760$ nm (Fig. 1f), the fringes almost disappear in the interior of the sample. In contrast, fringes on SiO$_2$ show more subdued variations with $\lambda_0$.

The observed fringe patterns on both WSe$_2$ and SiO$_2$ can be understood to be formed due to interference between photons collected by the objective from different paths. The two major paths relevant here are sketched in Fig. 1b. In path one (P$_1$), photons are directly reflected off the sample and collected by the objective. In path two (P$_2$), photons transfer into in-plane propagative modes and scatter back to photons after reaching the sample edge. Note that the in-plane modes contributing to P$_2$ are mainly those propagating normal to the sample edge (along the -$x$ direction in Fig. 1b). The photons scattered from in-plane modes propagating along other directions will not be efficiently collected by the top objective

(Supplementary Material). In addition to $P_1$ and $P_2$, there are also other possible photon paths, but they play less significant roles here (Supplementary Material). As will be discussed in detail below, these in-plane modes inside path $P_2$ on the sample side are a mixture of photons in air or $SiO_2$ and confined waveguide modes inside $WSe_2$. The latter are enabled by the sub-diffraction-limit aperture of the NSOM tip, which generates a wide range of in-plane momenta ($q$). The phase difference between collected photons from paths $P_1$ and $P_2$ is determined by the distance ($x$) between the tip and the sample edge. Therefore when scanning the tip away from the sample edge, one expects interference fringes with periodicities that equal to the wavelengths ($\lambda_p$) of the in-plane modes. The longest distance away from the edge that the fringes extend is determined by the propagation lengths ($L_p$) of these modes.

Based on the above picture, we are able to determine all the essential parameters of the in-plane modes by analyzing the fringe patterns. In Fig. 2a, we plot the line profiles extracted perpendicular to the fringes of $WSe_2$ and $SiO_2$ along the arrows marked in Fig. 1d ($\lambda_0$ = 850 nm). One can see that the fringe profile of $WSe_2$ consists of a number of oscillations that extend far away from the sample edge ($x$ = 0). In order to estimate $L_p$, we plot in Fig. 2b background-subtracted fringe profile of $WSe_2$ as well as calculated decay curves considering different $L_p$ (dashed curves, see Supplementary Material for details about the calculation). The best match to the data is obtained when $L_p \approx 3$ μm (black dashed curves). As detailed below, the oscillations on $WSe_2$ are due to a mixture of multiple in-plane modes, so our estimation produces an average $L_p$ over all the modes launched by the NSOM tip. Similar analysis is also performed on data profiles taken at other $\lambda_0$ (Fig. S4a). Thus-obtained $L_p(\lambda_0)$ is plotted in Fig. 2c as hollow squares, where one can see that $L_p$ increases rapidly with $\lambda_0$ above the A exciton wavelength ($\lambda_{ex} \approx 790$ nm). The trend of the rapid increase projects to a huge $L_p$ of tens of microns as $\lambda_0$ approaches 900 nm. Therefore, a much larger crystal with a scale over 100 μm is necessary to accurately determine $L_p$ at wavelengths close to 900 nm. Nevertheless, the large $L_p$ at these wavelengths is reflected by the weakly-damped fringes shown in the data image (Fig. 1b) and profiles (Fig. S4a) of our current sample with a practical size (Fig. S1b). In Fig. 2c, we also plot the theoretical calculation of the upper limit of $L_p$ at our $\lambda_0$ range (Supplementary Material), which captures well the trend of the experimental data points.

From Fig. 2a, we notice that the fringe profile of $WSe_2$ is complicated and appears to be a superposition of oscillations with different periodicities. In order to determine the periodicities accurately, we performed Fourier analysis on these real-space profiles. The outcomes are the $q$-space profiles (Fig. 2d,e), where the peaks correspond directly to the momenta ($q_p = 2\pi/\lambda_p$) of the in-plane modes. From Fig. 2d, one can see that the $q$-profile of $SiO_2$ has one broad peak close to $k_0 = 2\pi/\lambda_0$ indicating that they are mainly photons (see Supplementary Material for more discussions about bare substrate modes). The $q$-profile of $WSe_2$, on the other hand, shows multiple peaks (mark with arrows). The two peaks close to $k_0$ (black arrow) and $n_{sub}k_0$ (purple arrow, $n_{sub}$ = 1.46 is the refractive index of $SiO_2$) are air and substrate modes respectively. The field of the two modes are not confined inside

WSe$_2$ but extending to air or substrate [23]. In addition to the two modes, there are three more peaks above 1.46$k_0$ (blue, green and red arrows), which we label as various waveguide modes inside WSe$_2$ according to our dispersion analysis (see discussions below). Note that the mode at $q_p = 1.6k_0$ (blue arrow) merges with the adjacent substrate mode (purple arrow) and becomes a shoulder feature. The highest $q_p$ of these waveguide modes is close to 2.5$k_0$, corresponding to a $\lambda_p$ of about 350 nm. In order to study the $\lambda_0$-dependence of the in-plane modes inside WSe$_2$, in Fig. 2e we plot a complete set of $q$-profiles of WSe$_2$ at all $\lambda_0$. Here, one can see that all the modes (marked with arrows) evolve systematically with $\lambda_0$. As $\lambda_0$ decreases from 900 nm towards 800 nm, all the modes shift to larger $q$ indicating smaller mode wavelength ($\lambda_p$). When $\lambda_0$ reaches the exciton wavelength ($\lambda_{ex} \approx$ 790 nm) and below, the air and substrate modes (black and purple arrows) shift abruptly to lower $q$. The waveguide modes inside WSe$_2$ (blue, green and red arrows), on the other hand, become damped and are thus not clearly distinguishable. As $\lambda_0$ approaches 740 nm away from the A exciton, we observe a broad hump feature below $q = 2.0k_0$ (orange arrows in Fig. 2e and Fig. S4b). Such a hump feature are formed due to the merging of all the damped waveguide modes inside WSe$_2$ (see discussions below).

To gain insights into these modes, in Fig. 3a,b,d,e we plot the calculated $\lambda_0$- and $q$-dependent dispersion diagrams of the entire sample/substrate system. Because the aperture-type NSOM probe induce strong enhancement of both in-plane and out-of-plane fields, it can couple to both transverse electric (TE) and transverse magnetic (TM) modes [24,25]. Therefore we considered both TE and TM polarizations in our calculations. The dispersion diagrams in Fig. 3a,d are calculated with the realistic in-plane dielectric function of WSe$_2$ from literature [26], while those in Fig. 3b,e are calculated with artificial optical constants considering only the A exciton (Fig. S3). Details about the dispersion calculation and the optical constants of WSe$_2$ are introduced in the Supplementary Material. The bright curves shown in these dispersion plots correspond to various in-plane modes inside the system, among which we label the waveguide modes of WSe$_2$ as TE$_0$, TE$_1$, TE$_2$, TM$_0$ and TM$_1$ modes based on the calculated field distributions (Fig. 3c,f). There are also modes appearing at low $q$ regime close to $k_0$ and 1.46$k_0$ in the TM dispersion maps, which are air and substrate modes respectively. The data points overlaid on top the color maps in Fig. 3a,d are the $q$ positions for the air mode (black), substrate mode (purple) and various WSe$_2$ waveguide modes (red, blue and green) extracted from the Fourier-$q$ profiles in Fig. 2e. Note that we only show data points at selected $\lambda_0$ in Fig. 3a,d to avoid blocking the color map.

From Fig. 3 one can see that all the modes show distinct dispersion properties due to their coupling with the A excitons ($\lambda_{ex}$) in WSe$_2$ (white dashed lines in the dispersion plots). More specifically, when approaching excitons from higher excitation wavelengths ($\lambda_0 > \lambda_{ex}$) all the waveguide modes inside WSe$_2$ shift to higher $q$, which can be seen clearly from both experimental data points and calculated dispersion color plots with realistic optical constants (Fig. 3a,d). At lower excitation wavelengths ($\lambda_0 < \lambda_{ex}$), on the other hand, all the waveguide modes suffer from damping due to adjacent B excitons or band-edge absorption of WSe$_2$ (Fig. S3). As a result, lower branches of these waveguide modes ($\lambda_0 < \lambda_{ex}$) are not clearly resolved experimentally. Instead, they merge together as one broad resonance feature (Fig. 2e). Calculations considering only the A excitons (Fig. S2) show clearer the bottom branches (Fig. 3b,e). From the dispersion color plots (Fig. 3a,b,d,e), one can see clear anti-crossing behaviors above and below $\lambda_{ex}$ for all the waveguide modes, which are signatures of formation of exciton polaritons – hybrid modes between photons and excitons [27-30]. Recently, exciton polaritons in TMDCs drew a lot of research interests [31-34] and were observed by far-field spectroscopy with cavity coupling methods [33,34]. Our real-space nano-imaging data clearly capture the high-$\lambda_0$ (or low-energy) portion of the polariton dispersion of WSe$_2$ (Fig. 3a,d). In addition to the waveguide modes inside WSe$_2$, the air and substrate modes (black and purple data points in Fig. 3d) also couple to the WSe$_2$ excitons and thus become discontinuous above and below $\lambda_{ex}$. For the bare SiO$_2$/Si substrate, on the other hand, only straight photon modes between $k_0$ and $1.46k_0$ can be seen in their Fourier $q$ profiles (Fig. S4b) and also the calculated dispersion diagrams (Fig. S6). Note that the TE$_0$ and TM$_0$ modes of the 260-nm-thick WSe$_2$ (Fig. 3a,d) are not clearly resolved experimentally from the Fourier $q$ profiles (Fig. 2d,e) due to their narrow linewidths that are beyond the resolution limit of our NSOM probe (Supplementary Material).

Finally, we wish to discuss the dependence of the waveguide modes with the flake thickness. In Fig. 4a, we plot the near-field image of a 120-nm-thick WSe$_2$ flake taken at $\lambda_0 = 850$ nm. Following the same procedure as described above, we plot in Fig. 4b the real-space fringe profile (top panel) and the corresponding Fourier $q$-profile (bottom panel) of WSe$_2$. Unlike the 260-nm-thick flake discussed above, the current thinner sample has only two peaks at $q \approx 1.6k_0$ and $2.4k_0$ (grey and yellow arrows) above the far-field photon line ($q = k_0$). Similar analysis was also done on data taken from a 70-nm-thick flake (Fig. S7), where the two peaks shift down to $q \approx 1.5k_0$ and $2.0k_0$, respectively. As the sample gets even thinner down to a few nanometers (Fig. S8), the fringes appear to be sinusoid indicating that only single mode is involved. For example, the WSe$_2$ samples with a thickness of 8 nm and 4 nm have mode momenta of $q \approx 1.5k_0$ and $1.3k_0$, respectively (Supplementary Material).

To understand the observed thickness dependence, we plot in Fig. 4c,d the calculated thickness- and momentum-dependent color maps of both TE and TM modes at $\lambda_0 = 850$ nm, where the bright curves represent various modes in the system. The data points here are mode positions above $q = 1.46k_0$ extracted from mode analyses from multiple samples (Fig. 2, Fig. 4, Fig. S7, Fig. S8), and they are consistent with the calculated color plots. From Fig. 4c,d, one can see that all both the mode position and the number of modes (bright curves) evolve systematically with the flake thickness. For the 260-nm-thick flake, there are five waveguide modes above $1.46k_0$, among which the TE$_1$, TE$_2$ and TM$_1$ modes are clearly resolved by our Fourier analysis (Fig. 2d). While in the case of the WSe$_2$ flakes with a thickness of 120 nm and 70 nm, there are two waveguide modes (TE$_0$ and TM$_0$ modes) above $1.46k_0$ (Fig. 4c,d). The TE$_1$ mode here in the 120-nm-thick flake are below

1.46$k_0$, so it merged with the air and substrate modes and become together a broad low $q$ peak centered at $k_0$ in the Fourier $q$ profile. When the sample gets even thinner to a few nanometers, only the TE$_0$ mode exists, which is captured by our nano-optical imaging data (Fig. S8).

With the state-of-the-art near-field optical microscopy, we performed the first nano-optical imaging study of waveguide modes inside WSe$_2$. By recording and analyzing the interference fringes of these modes, we were able to extract both their wavelengths and propagation lengths. Such a surface interferometry method has been We found that the wavelengths of these modes are as low as 300 nm that are two to three times smaller than the excitation laser wavelengths and they can propagate over tens of microns below the A exciton energy. Moreover, we observed signatures of coupling between waveguide photon modes with A excitons in WSe$_2$. The coupling shift the waveguide modes towards higher momentum below the A exciton energy. As the excitation laser energy increases above the A exciton energy, all waveguide modes become strongly damped due to adjacent B excitons and band-edge absorption. Our work opens up a new regime for studying nanoscale light-matter interactions in TMDCs and thus provides guidelines for future applications of this class of materials in nanophotonics and optoelectronics.

**Acknowledgements**
This work was performed, in part, at the Center for Nanoscale Materials, a U.S. Department of Energy Office of Science User Facility under Contract No. DE-AC02-06CH11357. The work at UW was supported by the U.S. DOE Basic Energy Sciences, Materials Sciences and Engineering Division (DE-SC0008145 and SC0012509). The work at ORNL (JQY and DGM) was supported by the U.S. Department of Energy, Office of Science, Basic Energy Sciences, Materials Sciences and Engineering Division.

**Figures Captions**

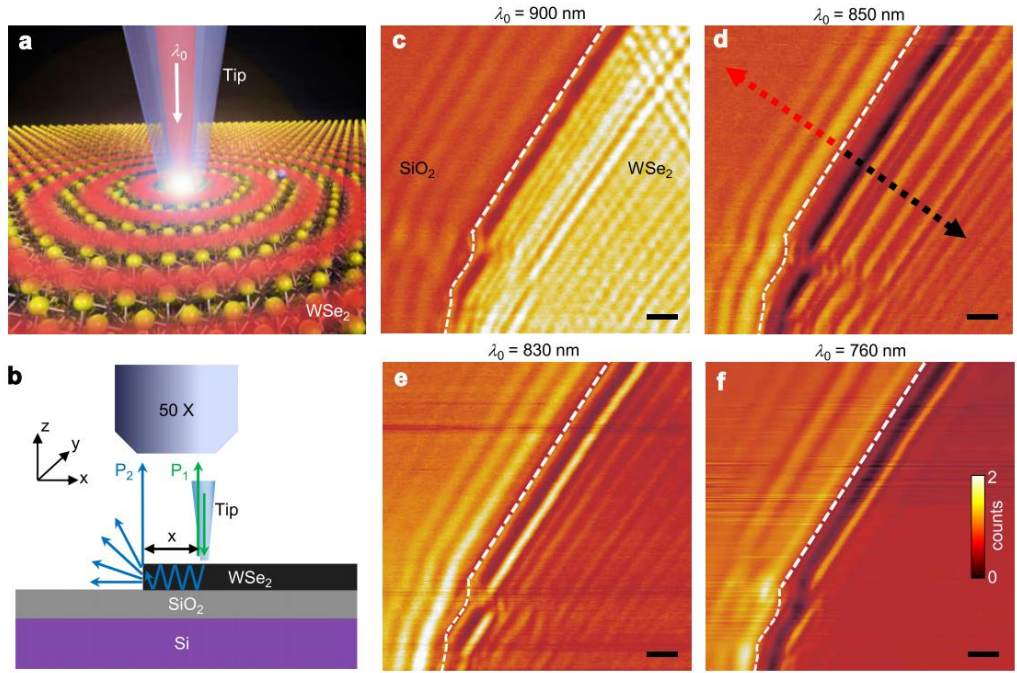

**Fig. 1. a**, Schematics of the near-field optical study of $WSe_2$. **b**, Illustration of the experimental setup and two major pathways ($P_1$ and $P_2$) that photons are collected by the top objective. **c-f**, Selected nano-optical imaging data of a $WSe_2$ flake (thickness = 260 nm) taken at various excitation wavelengths ($\lambda_0$). Here we plot the total counts of the photons collected by the objective normalized to that taken on the $SiO_2$/Si substrate. The white dashed lines mark the edges of the $WSe_2$ flake. The two arrows in **d** mark the directions along which we took line profiles across the fringes on both $WSe_2$ and $SiO_2$ as plotted in Fig. 2a. Scale bars in **c-f** represent 1 μm.

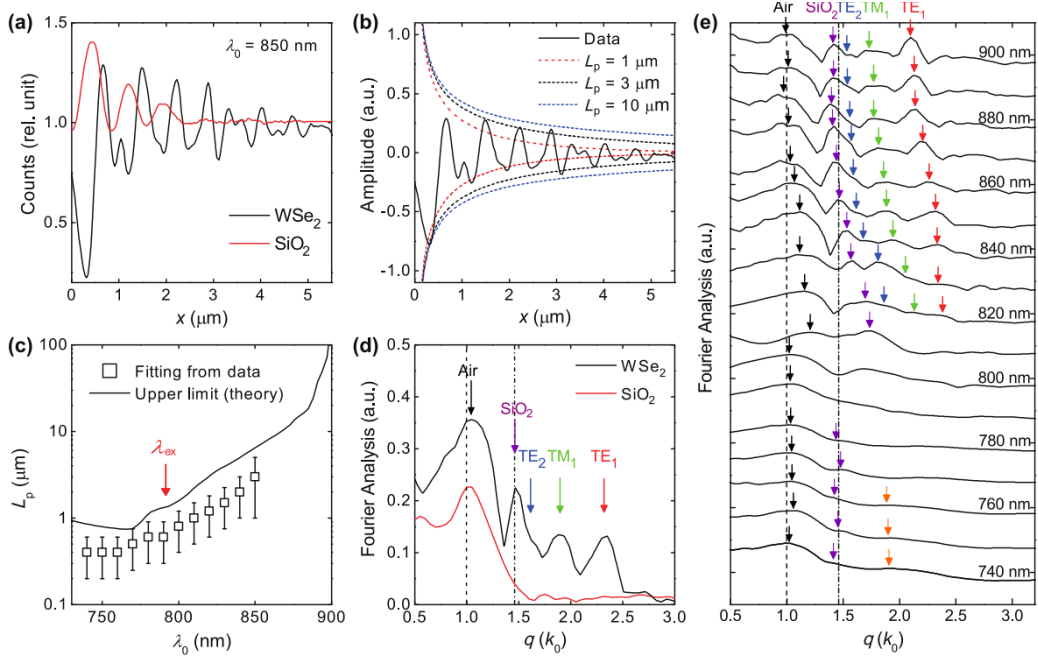

**Fig. 2. a**, Real-space fringe profiles of $WSe_2$ and $SiO_2$ at $\lambda_0 = 850$ nm taken directly from Fig. 1d along the black and red arrows, respectively. **b**, The same fringe profile of $WSe_2$ as **a** with background signal subtracted and the calculated decay curves by assuming different propagation lengths ($L_p$) (see Supplementary Materials). **c**, Estimated $L_p$ versus $\lambda_0$ from data fitting (hollow squares) and theoretical calculation of the upper limit of $L_p(\lambda_0)$ (black curve, see Supplementary Materials). The red arrow marks the A exciton wavelength ($\lambda_{ex}$). **d**, Momentum ($q$) profiles of $WSe_2$ and $SiO_2$ at $\lambda_0 = 850$ nm obtained by Fourier analysis of the real-space fringe profiles shown in **a**. **e**, The $q$-profiles of $WSe_2$ (black curves) at all $\lambda_0$ obtained by Fourier analysis of the real-space fringe profiles (Fig. S4a). The $q$-profiles are displaced vertically for clarity. The unit of the $q$ axis in **d,e** is the excitation far-field wavevector $k_0 = 2\pi/\lambda_0$. The vertical black dashed and dash-dotted lines in **d,e** mark the photon lines in air ($q = k_0$) and $SiO_2$ ($q = 1.46k_0$), respectively. The black and purple arrows in **d,e** mark the air and substrate modes. The blue, green and red arrows in **d,e** mark respectively the $TE_2$, $TM_1$ and $TE_1$ waveguide modes inside $WSe_2$. The orange arrows in **e** at low $\lambda_0$ regime mark a broad hump feature that occurs due to the merging of the damped waveguide modes.

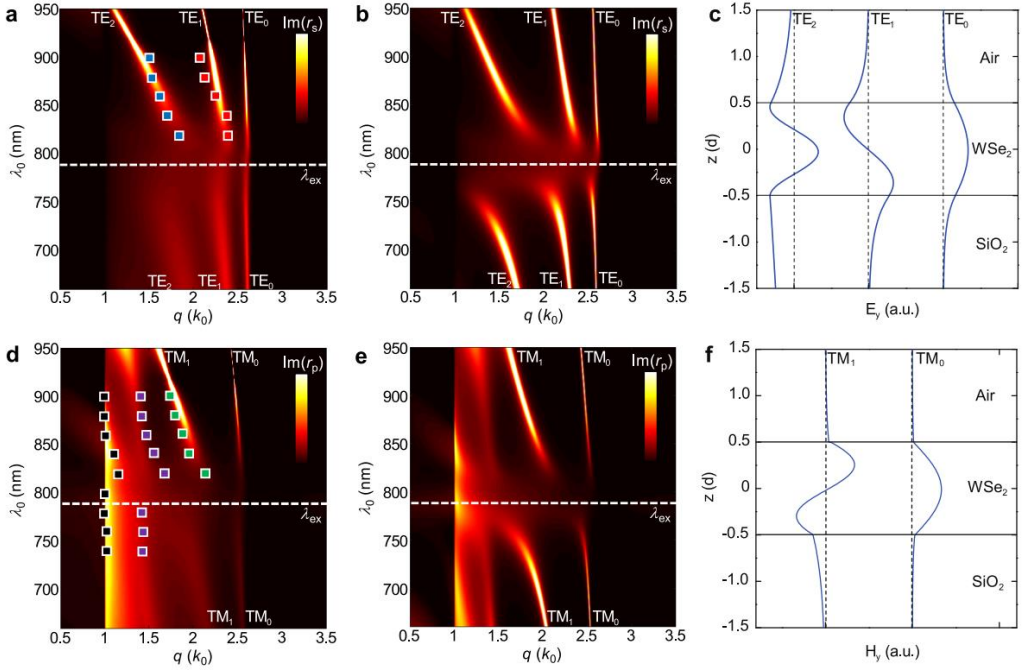

**Fig. 3. a,b**, Calculated color maps about the imaginary part of *s*-polarized reflection coefficient Im[$r_s(q, \lambda_0)$] of the WSe$_2$ sample (thickness = 260 nm) with both realistic and artificial optical constants (Fig. S2). **c**, Calculation of the *y*-direction electric field ($E_y$) of TE$_0$, TE$_1$ and TE$_2$ modes at $\lambda_0 = 900$ nm. **d,e**, Calculated color maps of the imaginary part of the *p*-polarized reflection coefficient Im[$r_p(q, \lambda_0)$] of the WSe$_2$ sample with both realistic and artificial optical constants. **f**, Calculation of the *y*-direction magnetic field ($H_y$) of TM$_0$ and TM$_1$ modes at $\lambda_0 = 900$ nm. The data points in **a,d** mark the peak positions extracted from the Fourier *q* profiles in Fig. 2e. To avoid blocking the color map, we only show data points at selected wavelengths here. The unit of the *q* axis in **a,b,d,e** is $k_0$. The bright curves in **a,b,d,e** above $q = 1.46k_0$ represent various waveguide modes inside WSe$_2$ as labeled in the figure. The horizontal dashed lines in **a,b,d,e** mark the wavelength for the A exciton of WSe$_2$ ($\lambda_{ex}$). The horizontal solid lines in **c,f** mark the air/WSe$_2$ and WSe$_2$/SiO$_2$ interfaces. The vertical dashed lines in **c,f** mark $H_y = 0$ and $E_y = 0$, respectively.

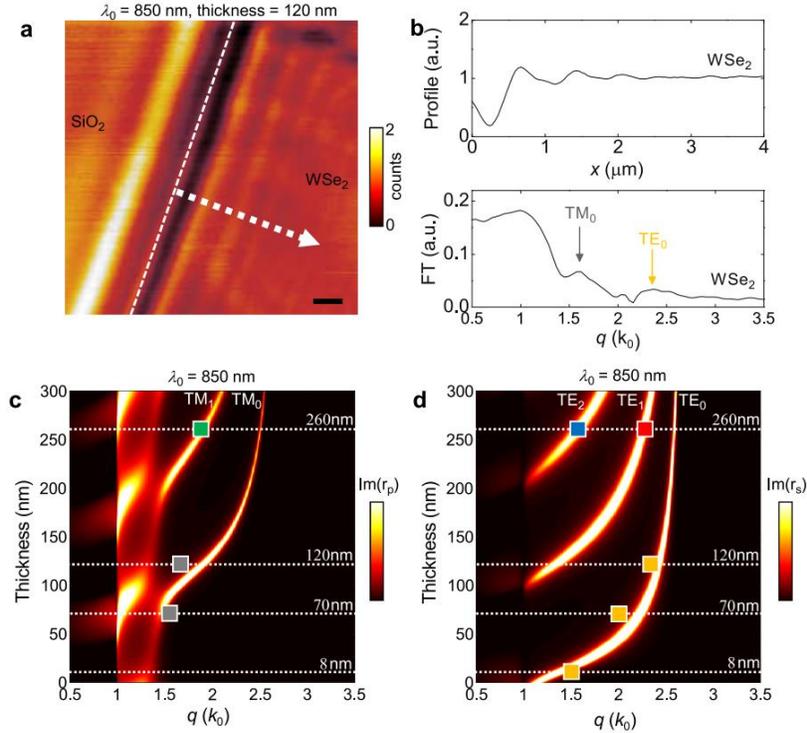

**Fig. 4. a**, Nano-optical imaging data of a $WSe_2$ thin flake (thickness = 120 nm) taken at $\lambda_0$ = 850 nm. The white dashed line marks the edge of the $WSe_2$ flake. Scale bar represents 700 nm. **b**, Real-space line profile (top panel) taken along the white arrow in **a** and the corresponding $q$-profile (bottom panel) obtained from Fourier analysis. The arrows here mark the positions of two waveguide modes above $q=1.46k_0$. **c,d**, Calculated thickness- and momentum-dependence color maps of $Im(r_p)$ and $Im(r_s)$ for TM and TE modes respectively. The data points overlaid on top the color maps mark the mode positions of $WSe_2$ waveguide modes extracted from mode analysis in Fig. 2, Fig. 4, Fig. S7 and Fig. S8.

# Supplemental Material for
# "Nano-optical imaging of WSe$_2$ waveguide modes revealing light-exciton interactions"

Z. Fei[1,2,*], M. E. Scott[3], D. J. Gosztola[1], J. J. Foley IV[1,4], J. Yan[5,6], D. G. Mandrus[5,6], H. Wen[7], P. Zhou[8], D. W. Zhang[8], Y. Sun[1], J. R. Guest[1], S. K. Gray[1], W. Bao[8], G. P. Wiederrecht[1], X. Xu[3,9]

[1]Center for Nanoscale Materials, Argonne National Laboratory, Lemont, Illinois 60439, USA
[2]Department of Physics and Astronomy, Iowa State University, Ames, Iowa 50011, USA
[3]Department of Physics, University of Washington, Seattle, Washington 98195, USA
[4]Department of Chemistry, William Paterson University, Wayne, New Jersey 07470, USA.
[5]Materials Science and Technology Division, Oak Ridge National Laboratory, Oak Ridge, Tennessee 37831, USA
[6]Department of Materials Science and Engineering, University of Tennessee, Knoxville, Tennessee 37996, USA
[7]Advanced Photon Source, Argonne National Laboratory, Lemont, IL 60439, USA
[8]State Key Laboratory of ASIC and System, Department of Microelectronics, Fudan University, Shanghai 200433, China
[9]Department of Materials Science and Engineering, University of Washington, Seattle, Washington 98195, USA

*Email: (Z.F.) zfei@iastate.edu.

**List of contents**

1. Experimental setup

2. Fringe formation

3. Optical constants of WSe$_2$

4. Estimation about the mode propagation length

5. Fourier analysis of fringe profiles

6. Dispersion calculation

7. Supplemental data

1. **Experimental setup**

   The aperture-type near-field optical microscope (NSOM) used in the current work is a commercial system from Nanonics. Inc. (MultiView 4000™). The NSOM tips are tapered optical fibers with metal coating and the aperture size of the NSOM tips is about 100 nm, nearly one order of magnitude smaller than the excitation wavelengths. These tips are mounted to a tapping-mode tuning fork with a tapping frequency of ~ 30 kHz. The excitation laser used in this work is a Ti: sapphire laser (Coherent, Mira 900) that is coupled to the NSOM with a multi-mode optical fiber. Throughout the experiments, we operate the laser in the continuous-wave mode. The spectral range of the laser is 700-950 nm and the spectral resolution is below 1 nm. We keep the laser power below 15 mW to avoid possible damage to the fiber and the NSOM tip. To collect and count photons, we used a long working distance objective (Nikon, N.A. = 0.45) and a silicon avalanche photodiode (PerkinElmer, SPCM-AQR series). Our $WSe_2$ samples are fabricated by mechanical exfoliation of high quality bulk crystals onto the standard $SiO_2$/Si substrates (Fig. S1a). The thickness of the $SiO_2$ layer is about 300 nm. All our measurements were performed under ambient conditions.

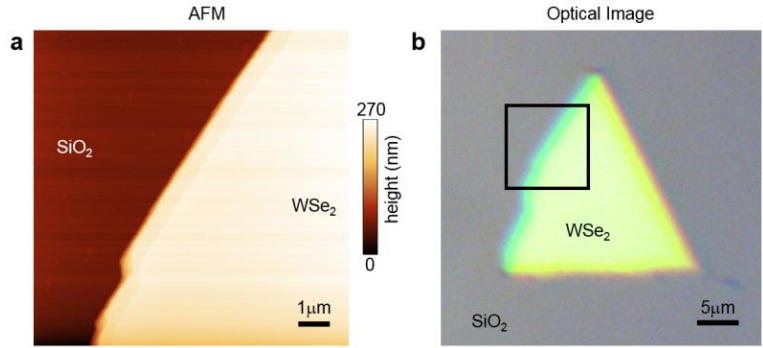

**Fig. S1. a,b** The AFM topography and optical images of the $WSe_2$ thin flake studied in Figs. 1 and 2 in the main text. The square in **b** marks the position where our AFM and near-field imaging data (Fig. 1 in the main text) were acquired.

2. **Fringe formation**

   In Fig. 1(b) in the main text, we illustrate two major paths that photons can be collected by the top objective. In path one ($P_1$), photons are directly reflected off the sample and collected by the objective. In path two ($P_2$), photons transfer into in-plane propagative modes and scatter back to photons after reaching the sample edge. In the latter case ($P_2$), the tip-launched waveguide modes propagate radially away from the tip, so they can be scattered into free-space photons from any positions along the sample edge. Nevertheless, our objective with a N.A. of 0.45 collects mainly the scattered photons from waveguide modes propagating normal to the edge. In order to explain that, we illustrate below (Fig. S2) the process of edge scattering where the $WSe_2$ edge is along the $y$ direction. Here, we consider in-plane modes ($x$-$y$ plane) with a momentum of $q_p$ propagating towards the edge of $WSe_2$ with an angle of $\phi$ with respect to the -x direction. They are scattered by the edge into free-space photons with a wavevector of $k_0 = \omega/c$. In order to satisfy the boundary condition of Maxwell's equations and momentum conservation, the y components of the wavevectors of waveguide modes and scattered photons should be equal: $k_y = q_p \sin(\phi)$. Therefore, we have $q_p \sin(\phi) = k_0 \sin(\theta)\sin(\psi)$. Consider that the objective can only collect photons with $\sin(\theta) <$ N.A. $= 0.45$ and that the $WSe_2$ waveguide modes have higher $q_p$ compared to photons in $SiO_2$ ($q_p > 1.46k_0$), we have $\sin(\phi) < 0.45(k_0/q_p)\sin(\psi) < 0.3\sin(\psi) < 0.3$. Therefore, we have at least $\phi < 17°$. Note that the above inequality analysis is far from stringent, more careful analysis will lead to an average $\phi$ much closer to zero. In addition, compared to waveguide modes with normal incidence towards the edge ($\phi = 0$), those with finite $\phi$ will suffer from higher propagation loss and larger reflection coefficient. Therefore, we consider mainly the photons scattered from normal-incident modes. Similar conclusion was also obtained both experimentally and theoretically from a previous study [1], where these authors studied edge scattering of surface plasmon polaritons and found that N.A. is a critical parameter governing the photon collection process.

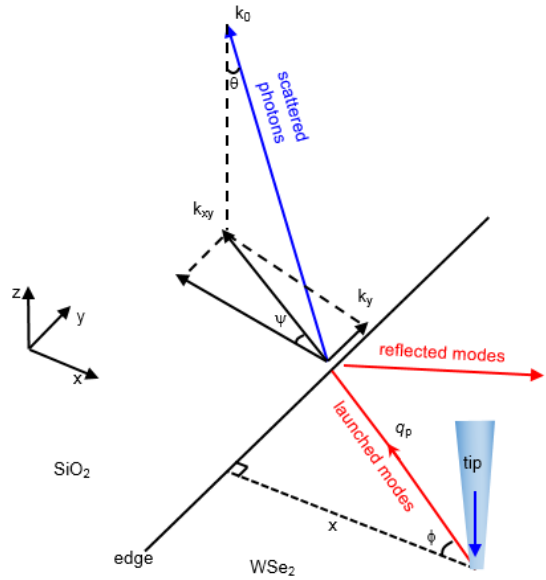

**Fig. S2.** Illustration of the photon path $P_2$ where the tip-launched in-plane modes are scattered by the edges into free-space photons.

In addition to $P_2$, there are also other possible photon paths. For example, in path $P_3$, the photons first transfer into in-plane modes and then reflected back after reaching the sample edge. The reflected modes can be scattered out by the AFM tip and collected by the top objective. This path has been extensively discussed in previous nano-infrared imaging studies of graphene plasmons [2] and hexagonal boron nitride (hBN) polaritons [3] by the scattering-type NSOM. In the current work, $P_3$ plays a less significant role compared to $P_2$ for the following reasons. First of all, the momenta (q) of the modes involved in our experiments is much closer to free-space photon wavevector ($k_0$). As a result only a small portion of the modes are reflected back by the sample edge. For example, for the modes with $q \sim 2k_0$ (e.g. $TE_1$ or $TM_1$ modes plotted in Figure 3), the reflection coefficient is about 1/3, so the energy of reflected mode is only about 11% of the incident mode energy. In addition, the round-trip propagation of the modes in $P_3$ also costs more damping compared to $P_2$ due to the longer traveling distance. Furthermore, the size of the fiber tip with thick coating in the aperture-type NSOM is hundreds of nanometers in diameter, which is much bigger than that of the scattering-type NSOM. Such a large tip leads to an inefficient and incoherent scattering of the modes inside the sample. Therefore, we didn't consider $P_3$ in our data analysis.

### 3. Optical constants of WSe$_2$

The *ab*-plane dielectric constants of WSe$_2$ used in our calculations for Fig. 3(a) and 3(d) in the main text are from Ref. [4] as replotted in Fig. S3, where one can see clearly the A and B excitons of WSe$_2$. Finite $\varepsilon_2$ at energies higher than the A exciton energy, which is due to the adjacent B exciton and band edge absorption, is responsible for the high mode damping at $\lambda_0 < \lambda_{ex}$. The dashed lines are artificial *ab*-plane optical constants we constructed by using one Lorentzian oscillator for the A exciton. Dispersion calculations based on the artificial dielectric function (Fig. 3b and 3e in the main text) clearly reveal the low-$\lambda_0$ branches for various waveguide modes. The *c*-axis dielectric constant of WSe$_2$ is set to be 7.2 according to a recent experimental study [5]. As reported in Ref. [6], the reflectance spectrum when the electric field is parallel to the *c*-axis is relatively flat at our $\lambda_0$ region, therefore we didn't consider $\lambda_0$-dependence for the *c*-axis dielectric constant in our calculations.

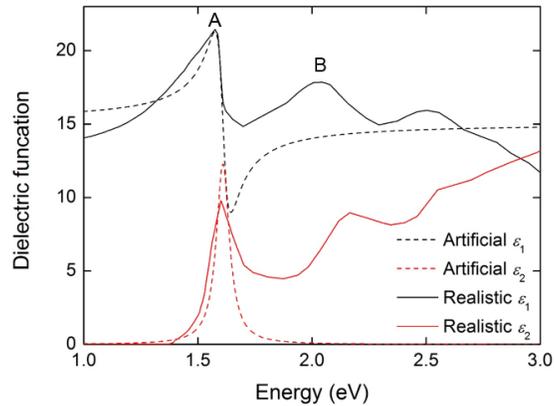

**Fig. S3.** The *ab*-plane dielectric function of WSe$_2$ for calculation of the dispersion diagrams of the waveguide modes. The solid lines are realistic optical constants adopted from Ref. [4]. The dashed lines are artificial optical constants we constructed by using one Lorentzian oscillator for the A exciton.

## 4. Estimation about the mode propagation length

The waveguide modes launched by the NSOM tip can be approximated as cylinder waves and the wave function can be written as $Ax^{-1/2}e^{i(qx-\omega t)}$. Here, $q = q_1 + iq_2$ is the complex in-plane momentum of a particular mode. Therefore, the amplitude of the wave decays as: $Ax^{-1/2}e^{-x/(2L_p)}$, where the propagation length $L_p$ equals to $(2q_2)^{-1}$. Based on the above decay function, we produce the mode decay curves in Fig. 2b (dashed curves) and extract $L_p$ by matching the decay curves with the fringe profile. Thus estimated $L_p$ is an average value over all the in-plane modes launched by the NSOM tip. For an anisotropic material like WSe$_2$ with an $ab$-plane dielectric function of $\varepsilon_{ab} = \varepsilon_1 + i\varepsilon_2$ (Fig. S3) and an $c$-axis dielectric constant of $\varepsilon_c \approx 7.2$ [5], we have $q = \sqrt{\varepsilon_c k_0^2 - (\varepsilon_c/\varepsilon_{ab})k_z^2}$. Within the most relevant $\lambda_0$ regime: 800 – 900 nm, where $\varepsilon_1 \gg \varepsilon_c \gg \varepsilon_2$, we have $L_p \approx (\varepsilon_1/\varepsilon_2)*[q_1/(\varepsilon_c k_0^2 - q_1^2)]$. Based on this formula, we know that $L_p$ increases with $q_1$. According to the Fourier analysis (Fig. 2e) and the calculated dispersion diagrams (Fig. 3), the highest $q_1$ of all the waveguide modes is close to $2.5k_0$, which places an upper limit for $L_p$ of about $0.4\lambda_0\varepsilon_1/\varepsilon_2$. In Fig. 2c we plot a more accurate estimation about this upper limit with numerical methods (black curve), which matches quite well this simple analytic formula.

## 5. Fourier analysis of fringe profiles

In order to extract the periodicities of the waveguide modes, we performed discrete Fourier transform of the fringes profiles shown in Fig. 2a and Fig. S4a. The resolution of the thus obtained Fourier $q$-profile is about 1.14 μm$^{-1}$ (~ 0.15$k_0$ at 850 nm) limited by the finite scanning range (~ 5.5 μm in the sample side), which is sufficient to resolve most of the modes except TE$_0$ and TM$_0$ modes for the 260-nm-thick WSe$_2$ sample. Both of these two modes have a full width at half maximum of merely ~0.1$k_0$ according to our dispersion calculation (Fig. 3), therefore they are not clearly resolved by the Fourier analysis. These two modes become observable in the case of thinner WSe$_2$ flakes (Fig. 4 and Fig. S7), where their positions move to smaller $q$ and their widths increase significantly.

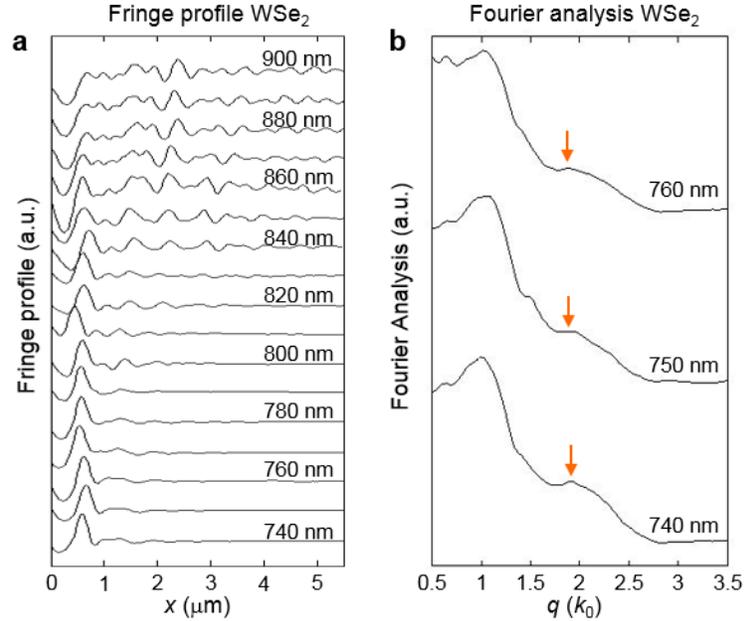

**Fig. S4**. **a**, Real-space line profiles across the fringes of WSe$_2$ taken perpendicular to the sample edge (along the black arrows in Fig. 1d) from data images at various $\lambda_0$ including those shown Fig. 1 in the main text. **b**, Momentum ($q$) space profiles (black curves) obtained by Fourier analysis of the real-space fringe profiles of WSe$_2$ at $\lambda_0 = 740$, 750 and 760 nm. All the real-space and $q$-space profiles are re-scaled and displaced vertically for clarity. The unit of the $q$ axis in **b** is $k_0$. The orange arrows here mark a broad hump feature close to $q = 2k_0$. Such hump features occur due to the merging of the damped TE$_1$, TE$_2$ and TM$_1$ waveguide modes in WSe$_2$.

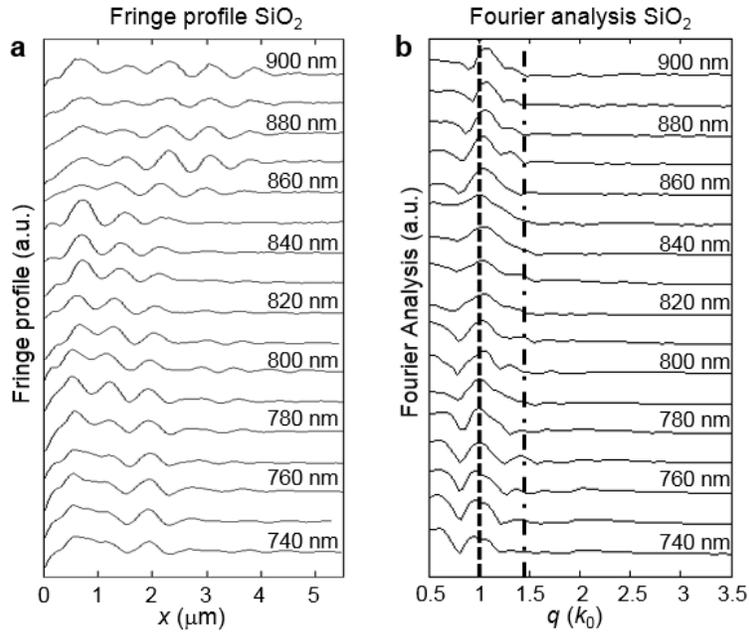

**Fig. S5**. **a**, Real-space line profiles across the fringes of the bare $SiO_2$/Si substrate taken perpendicular to the sample edge (along the red arrows in Fig. 1d in the main text) from data images at various $\lambda_0$ including those shown Fig. 1. **c**, Momentum ($q$) space profiles (black curves) obtained by Fourier analysis of the real-space fringe profiles of $SiO_2$ shown in **a**. The vertical black dashed lines mark the light line in air ($q = k_0$) and $SiO_2$ ($q = 1.46k_0$) respectively, where $k_0$ is the free-space wavevector. All the real-space and $q$-space profiles are re-scaled and displaced vertically for clarity. The unit of the $q$ axis in **c** is $k_0$.

## 6. Dispersion calculation

The dispersion diagrams shown in Fig. 3 of the main text are obtained by evaluating numerically the imaginary part of the reflection coefficients: $Im[r_s(q, \lambda_0)]$ for TE modes (*s*-polarized) and $Im[r_p(q, \lambda_0)]$ for TM modes (*p*-polarized) using the transfer matrix method. This method has been applied extensively in studying dispersion properties of other interesting modes, such as surface plasmons in graphene [7] and surface phonon polaritons in hexagonal boron nitride [3]. The field distribution plots in Fig. 3(c) and 3(f) in the main text are obtained by matching boundary conditions in Maxwell's wave equations.

## 7. Discussion about bare substrate modes

In the main text, we discuss mainly the modes on the sample side. Here we wish to discuss briefly the bare substrate modes based on both data analysis (Fig. S5) and dispersion calculation (Fig. S6). As shown in Fig. S5b, the dominant peak at the Fourier-$q$ profile of the substrate is sitting at $k_0$ indicating that they are in-plane photons propagating in air above the $SiO_2$ surface. In addition, we also see additional peak features both below and above $k_0$. The modes between $k_0$ and $1.46k_0$ are waveguide modes of $SiO_2$ as shown in the calculated dispersion diagram in Fig. S6. Because silicon has a larger index of refraction than $SiO_2$, these waveguide modes are not strictly confined inside $SiO_2$ and their energy are losing into silicon. The peaks inside the light line close to $0.5k_0$ at low $\lambda_0$ regime in Fig. S5b are due to constructive interference of incident far-field light, which is also captured by the dispersion plot in Fig. S5b.

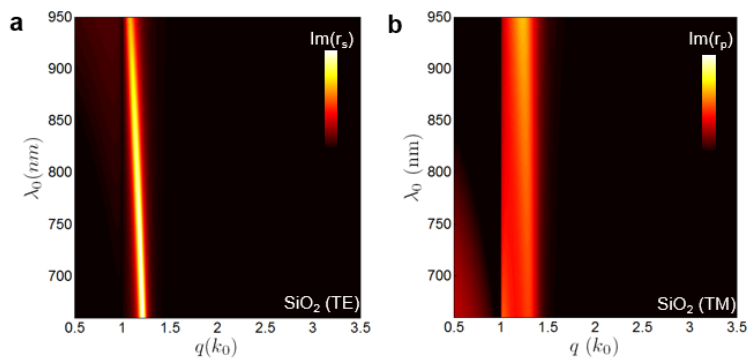

**Fig. S6**. Calculated dispersion diagrams of bare substrate with a 300-nm-thick $SiO_2$ layer

on Si, where we plot the imaginary parts of the reflection coefficient Im($r_s$) and Im($r_p$) for TE (**a**) and TM (**b**) polarizations, respectively. The bright lines sitting between $q = k_0$ and $q = 1.46k_0$ account for the waveguide photon modes in $SiO_2$.

## 8. Supplemental data

In Figs. S7 and S8, we present nano-optical imaging datasets of two additional $WSe_2$ samples. The data and analysis shown in Fig. S7 are for a $WSe_2$ thin flake with a thickness of about 70 nm. From the AFM topography image (Fig. S7a), we found a step feature close to the edge of the flake. Due to the step feature, there are standing wave pattern appearing inside the gap between the edge and the step, which can be seen clearly in the nano-optical image taken at $\lambda_0 = 850$ nm (Fig. S7b). In order to avoid the step-induced standing wave pattern, we took the line profile starting from the step towards the interior of the sample (white dashed arrows). Thus-obtained profile is plotted in Fig. S7c and the corresponding Fourier $q$ profile is given in Fig. S7d. From the $q$ profile, we found two prominent peaks close to $q \approx 1.5k_0$ and $q \approx 2.0k_0$ ($k_0 = 2\pi/\lambda_0$) respectively, which are associated with the $TM_0$ and $TE_0$ waveguide modes of $WSe_2$.

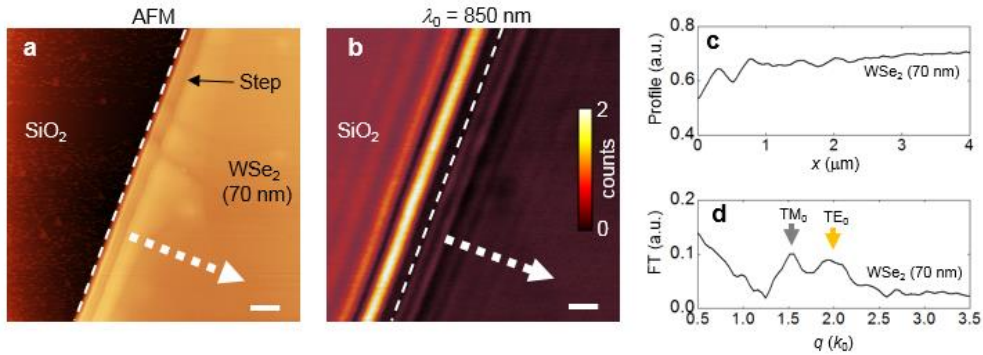

**Fig. S7. a,b,** The AFM topography and nano-optical images of a $WSe_2$ thin flake (thickness = 70 nm) taken at $\lambda_0 = 850$ nm. The white dashed line marks the edge of the $WSe_2$ flake. Scale bar represents 1 μm. **c,** The real-space line profile taken along the white arrow in **a**. **d,** the corresponding $q$-profile obtained from Fourier analysis of the profile in **c**. The arrows here mark the positions of two waveguide modes.

In Fig. S8, we present the data and analysis for a few-layer $WSe_2$ sample. From the AFM topography image (Fig. S8a), we found that the bottom part of the sample has a thickness of ~ 4 nm (~5 atomic layers) and the top part of the sample has a thickness of ~ 8 nm (~10 atomic layers). From the nano-optical image ($\lambda_0 = 850$ nm), we observed clear single-period fringes in both parts of the sample. We believe that the mechanism of the fringe formation here in few-layer $WSe_2$ is similar to that of bulk flakes. In both cases, the fringes are formed due to the interference between photons directly reflected off the sample (P1) and those scattered by the edges (P2). In principle, the scattering efficiency of photons off the edges of atomic layers of $WSe_2$ should be much weaker. Nevertheless, we noticed that there is some folded feature close to the edge of the few-layer $WSe_2$ sample (Fig. S8a), which could significantly enhance the scattering efficiency. The fringe periods for the 8-nm-thick and 4-nm-thick $WSe_2$ layers measured directly from Fig. S8b are about 570 nm and 650 nm corresponding to in-plane momenta of $1.5k_0$ and $1.3k_0$, respectively. Based on the calculated thickness dependence plot (Fig. 4c in the main text), we assign the observed modes here in few-layer $WSe_2$ to $TE_0$ modes.

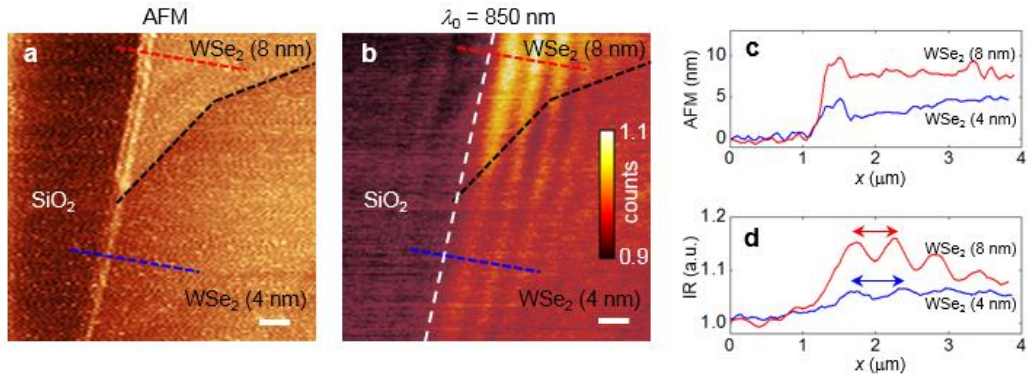

**Fig. S8. a**, The AFM topography image of a few-layer $WSe_2$ sample. **b**, Simultaneously-acquired nano-optical imaging data with an excitation laser wavelength of $\lambda_0 = 850$ nm. The white dashed line marks the edge of the $WSe_2$ flake. Scale bars in **a**,**b** represents 700 nm. **c,** The topography line profiles of the few-layer $WSe_2$ taken along the red and blue dashed lines in **a**. **d,** The nano-optical line profiles of the few-layer $WSe_2$ taken along the red and blue dashed lines in **b**. The red and blue arrows mark the sizes of the period of the signal oscillations.

**References for Supplementary Material**